\documentclass{ws-p8-50x6-00}

\begin{document}

\title{Modern Shell-model Calculations}

\author{A. Covello, L. Coraggio, A. Gargano, N. Itaco}

\address{Dipartimento di Scienze Fisiche, Universit\`a di Napoli Federico II, \\
and Istituto Nazionale di Fisica Nucleare, \\
Complesso Universitario di Monte S. Angelo, Via Cintia, 80126 Napoli, Italy \\ }

\author{T. T. S. Kuo}

\address{Department of Physics, SUNY, Stony Brook, New York 11794, USA}


\maketitle

\abstracts {We have performed shell-model calculations starting from the CD-Bonn free nucleon-nucleon ($NN$) potential and making use of a new approach for the
calculation of the effective interaction. This consists in deriving from the $NN$ potential a low-momentum effective potential $V_{low-k}$, defined within a given cut-off momentum $\Lambda$. Being a smooth potential it can be used directly in nuclear structure calculations. To assess the practical value of this approach, we have also performed calculations within the
framework of the usual Brueckner $G$-matrix formalism. Comparison of the calculated results between them and with the experimental data shows that the $V_{low-k}$ approach represents an advantageous alternative to the use of the $G$-matrix. }

\section{Introduction}
The shell model is the basic framework for nuclear structure calculations
in terms of nucleons. To make it a truly microscopic approach  
demands, however, that the model-space effective interaction $V_{\rm eff}$
be derived from the nucleon-nucleon ($NN$) interaction. This has long been indeed
a topical problem in nuclear structure theory, the first steps in this direction
dating back to the mid 1960s.\cite{kuo66} The two basic ingredients which come into play in what have become known as ``realistic" shell-model calculations are the free $NN$ potential and the many-body methods for deriving the effective interaction in the chosen model space. 
>From the early 1980s on there has been much progress in this field.  On the one hand, high-quality $NN$ potentials have been constructed which reproduce quite accurately the $NN$ scattering data and are suitable for application in nuclear structure.
On the other hand, an accurate calculation of the Brueckner $G$ matrix has become feasible while the so-called folded-diagram series for the effective interaction  
$V_{\rm eff}$ can be summed up to all orders using iterative methods. These
improvements have brought about a revival of interest in realistic shell-model calculations.

In this context, the main question is how accurate a description of nuclear structure properties can be provided by effective interactions derived from the free $NN$ potential. In recent years, we have concentrated our efforts on this problem studying a number of nuclei around doubly magic $^{100}$Sn, $^{132}$Sn, and $^{208}$Pb.$^{2-9}$
Our attention has been focused on nuclei with
few valence particles or holes since they provide the best testing ground
for the basic ingredients of shell-model calculations, especially as regards
the matrix elements of the two-body effective interaction. 
The results of our calculations have turned out to be in remarkably good
agreement with experiment for all the nuclei considered (about 20), providing evidence
that realistic effective interactions are able to describe with quantitative
accuracy the spectroscopic properties of complex nuclei. A brief survey of the present status of realistic shell-model calculations based on this comprehensive study has been given in Ref. 10. 

While we have also performed calculations starting from different $NN$ 
potentials,\cite{andr96,cov97,cov99} all the results summarized in Ref. 10 have been obtained by making use of effective interactions derived from the Bonn-A free $NN$ potential\cite{machl89} by means of a $G$-matrix formalism including
renormalizations from both core polarizations and folded diagrams.  

The use of the $G$ matrix has long been proved to be a valuable tool to overcome the difficulty posed by the strong repulsive core contained in all modern $NN$ potentials. 
As mentioned above, an accurate calculation of it can be carried out which is based on
the Tsai-Kuo method.\cite{tsai72} It should be recalled, however, that the $G$ matrix is model-space dependent, which makes its actual calculation rather involved. 
Very recently, a completely new approach has been proposed\cite{kuochina} which consists in deriving from the $NN$ potential a low-momentum effective potential $V_{low-k}$, defined within a given cut-off momentum $\Lambda$. This is a smooth potential which may be used directly in nuclear structure calculations.

The main aim of this paper is to report on realistic shell-model calculations performed by using as initial input the CD-Bonn potential\cite{machl01} and deriving the effective interaction by means of both the $G$-matrix and the $V_{low-k}$ formalisms. Our calculations concern the $^{132}$Sn neighbors with two, three, and four valence protons,
namely $^{134}$Te, $^{135}$I, and $^{136}$Xe.

Our presentation is organized as follows. In Sect. 2 we give an outline of our calculations. In Sect. 3 we present the results obtained by using the $V_{low-k}$ as well as the $G$-matrix and compare them with the experimental data. Some concluding remarks are given in Sect. 4.

\section{Outline of calculations}

As already mentioned in the Introduction, we present in this paper the results of calculations performed by using a new approach to the derivation of $V_{\rm eff}$
as well as those obtained by the usual $G$-matrix formalism.

A description of our procedure for deriving $V_{\rm eff}$ within the framework of a $G$-matrix folded-diagram formalism can be found in Refs. 15 and 16. We only outline here the essentials of the method to point out the main differences between this approach  
and that based on the use of $V_{low-k}$. 

The $G$ matrix is defined\cite{krenc76} by the integral equation

\begin{equation}
G(\omega)=V+VQ_2 \frac{1}{\omega-Q_2TQ_2}Q_2G(\omega),
\end{equation}

\noindent
where $V$ represents the $NN$ potential, $T$ denotes the two-nucleon kinetic
energy, and $\omega$ is an energy variable (the so-called starting energy).
The two-body Pauli exclusion operator $Q_2$ prevents double counting, namely
the intermediate states allowed for $G$ must be outside of the chosen model
space. Thus the Pauli operator $Q_2$ is dependent on the model space, and
so is the  $G$ matrix. The operator $Q_2$ is specified, as discussed in Ref. 17,  by a set of three numbers ($n_1, n_2, n_3$), each representing a shell-model orbital. We employ a matrix inversion method\cite{tsai72,krenc76}
to calculate the above $G$ matrix. This method gives the solution of Eq.(1) as
\begin{equation}
G=G_F +\Delta G.
\end{equation}
The ``free" $G_F$ matrix is
\begin{equation}
G_F=V+V\frac{1}{e}G_F,
\end{equation}
and the Pauli correction term $\Delta G$ is given by
\begin{equation}
\Delta G=-G_F\frac{1}{e}P_2\frac{1}{P_2(\frac{1}{e}+\frac{1}{e}G_F\frac{1}{e}
)P_2}P_2
\frac{1}{e}G_F,
\end{equation}
with $e\equiv (\omega - T)$ and $P_2=1-Q_2$.
The calculation of $G_F$ is straightforward, as it does not contain the Pauli
projection operator. Then $\Delta G$ can be calculated by performing some matrix
operations in the model space $P_2$.

Using the above $G$ matrix we can calculate $V_{\rm eff}$. This can be written schematically in operator form\cite{kuo80} as
\begin{equation}
V_{eff} = \hat{Q} - \hat{Q'} \int \hat{Q} + \hat{Q'} \int \hat{Q} \int
\hat{Q} - \hat{Q'} \int \hat{Q} \int \hat{Q} \int \hat{Q} + ~...~~,
\end{equation}

\noindent
where $\hat{Q}$ (referred to as  $\hat{Q}$-box) is a vertex function composed of irreducible linked diagrams, and the integral sign represents a
generalized folding operation. $\hat{Q'}$ is obtained from $\hat{Q}$ by removing terms first order in $V$. In our calculations we take the
$\hat{Q}$-box to be composed of $G$-matrix diagrams through second order.
After the $\hat{Q}$-box is calculated, $V_{\rm eff}$ is
obtained by summing up the folded-diagram series of Eq.(5) to all orders
by means of the Lee-Suzuki iteration method.\cite{suzuki80}
This last step can be performed in an essentially exact way for a given
$\hat{Q}$-box.

A detailed description of the derivation of  $V_{low-k}$ as well as a discussion of its main features can be found in Refs. 13 and 20.
The basic idea of this approach is to reduce a realistic $NN$ potential $V_{NN}$ to an effective low-momentum potential $V_{low-k}$ by integrating out the high-momentum modes. The $V_{low-k}$ constructed in this way is confined within a cut-off momentum $\Lambda$, and reproduces exactly the deuteron binding energy and the low-energy phase shifts (up to $E_{Lab} = 2 \hbar^2 \Lambda^2/M$) of $V_{NN}$.
It has the desirable feature of being a smooth potential and as such it is suitable to be used directly in shell-model calculations. The $V_{\rm eff}$ is derived by following the same procedure as that outlined above, except that the $G$ matrix is replaced by $V_{low-k}$. More explicitly, we first calculate the 
$\hat{Q}$-box including diagrams up to second order in $V_{low-k}$ and then obtain $V_{\rm eff}$ by summing up the folded-diagram series using the Lee-Suzuki iteration method. 

In our study of $^{134}$Te, $^{135}$I, and $^{136}$Xe we assume that $^{132}$Sn is a closed core and let the valence protons occupy the five single-particle (s.p.) orbits
$0g_{7/2}$, $1d_{5/2}$, $1d_{3/2}$, $2s_{1/2}$, and $h_{11/2}$. As regards the energy
spacings between the five s.p. levels, we take three of them from the experimental spectrum\cite{sanchez98} of $^{133}$Sb. As for the $s_{1/2}$ state, we adopt the value
$\epsilon_{1/2}$=2.8 MeV, which reproduces the experimental energy of the ${1 \over 2}^+$ level at 2.15 MeV in $^{137}$Cs. For the shell-model oscillator parameter $\hbar \omega$  we use the value 7.88 MeV, as obtained from the expression $\hbar \omega=45A^{-1/3}-25A^{-2/3}$ for $A=132$.

In the $G$-matrix calculations we have fixed ($n_1, n_2, n_3$)=(16,28,55) for the neutron orbits, and ($n_1, n_2, n_3$)=(11,21,55) for the proton orbits. As regards
the calculations using $V_{low-k}$ as the input interaction, an important question is how to choose the momentum cut-off $\Lambda$. According to the considerations made in Refs. 13 and 20, we have used in our calculation the value $\Lambda=2.05$ fm$^{-1}$. We have verified that the results are practically insensitive to changes in $\Lambda$
within reasonable limits, namely 1.9-2.2 fm$^{-1}$.

In both the $G$-matrix and the $V_{low-k}$ approaches the effective interaction has been derived from the CD-Bonn free $NN$ potential which fits\cite{machl01} the world $NN$ data below 350 MeV with a $\chi^2$/datum of 1.02.
Once the effective interaction has been derived, the shell-model calculations are carried out employing the OXBASH code.\cite{oxbash}
\begin{figure}[h] 
\begin{center}
\epsfxsize=9cm
\epsfbox{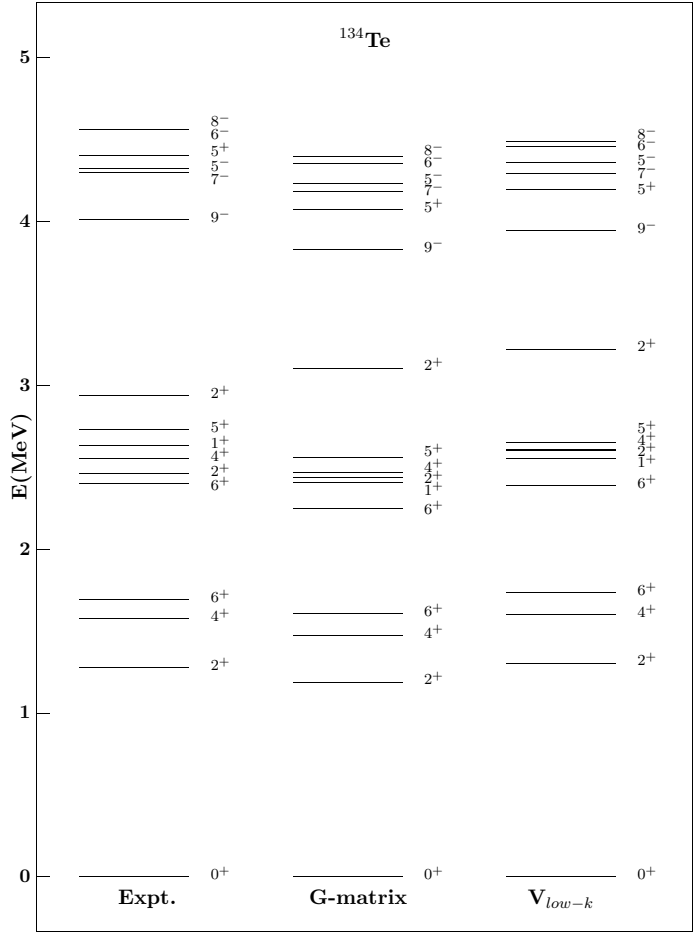}
\end{center}
\caption{Experimental and calculated spectra of $^{134}$Te.}
\end{figure}
\section{Results}

We present here the results of our realistic shell-model calculations for the $N=82$ isotones $^{134}$Te, $^{135}$I, and $^{136}$Xe. As already pointed out in the preceding sections, these results have all been obtained starting from the CD-Bonn $NN$ potential.
In Figs. 1-3 we report the experimental$^{23-25}$ and calculated spectra of these three nuclei. The latter have been obtained by using as input interaction the usual Brueckner $G$-matrix and the low-momentum $NN$ potential $V_{low-k}$. 

The spectra of $^{134}$Te and $^{135}$I obtained by using the Bonn-A $NN$ potential\cite{machl89} and the $G$-matrix formalism have been presented in a previous \hbox{paper,\cite{andr97}} where a detailed comparison between theory and experiment is made. While we refer
\begin{figure}[h] 
\begin{center}
\epsfxsize=9cm
\epsfbox{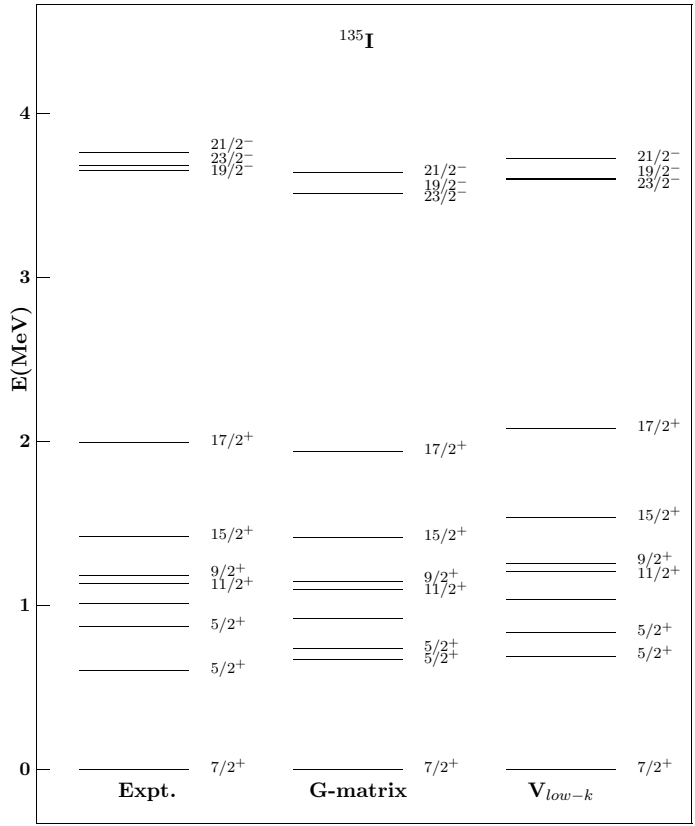}
\end{center}
\caption{Experimental and calculated spectra of $^{135}$I}
\end{figure}
the reader for details to the above paper, we emphasize here the very good agreement with experiment of both the calculated spectra. A measure of the overall agreement between the calculated and experimental energy levels is provided by the $rms$ deviation $\sigma$.\cite{sigma} For the spectra obtained using $V_{low-k}$ the values of $\sigma$ are 110, 73, and 111 keV for $^{134}$Te, $^{135}$I, and $^{136}$Xe, respectively, while for the $G$-matrix calculations
they are 163, 101, and 154 keV. 
This comparison shows that the $V_{low-k}$ results are rather similar to, and actually slightly better than, the $G$-matrix ones.
 \begin{figure} 
\begin{center}
\epsfxsize=9cm
\epsfbox{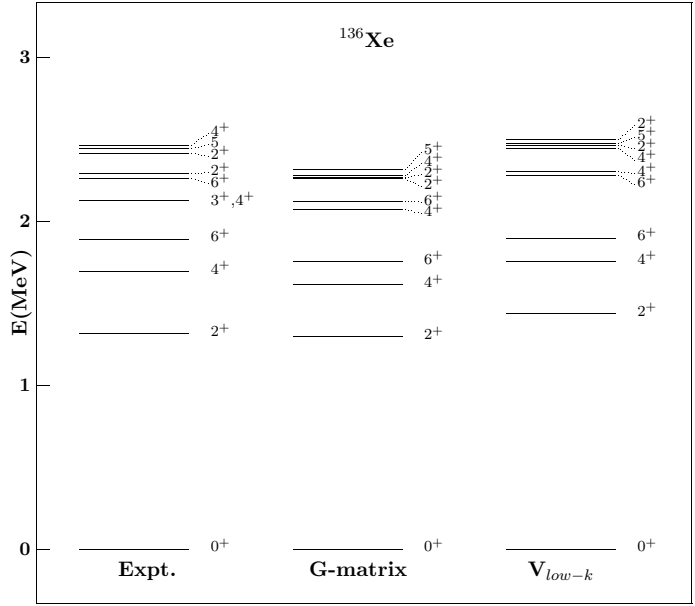}
\end{center}
\caption{Experimental and calculated spectra of $^{136}$Xe}
\end{figure}
\section {Concluding remarks}

In this paper, we have presented some results of realistic shell-model calculations
which have been performed within the framework of a new approach to the derivation of the effective interaction. This is based on the use of a low-momentum $NN$ potential
$V_{low-k}$ obtained by integrating out the high-momentum part of the original $V_{NN}$ potential containing a strong repulsive core. This $V_{low-k}$ is a smooth potential and can be used directly in nuclear structure calculations. To assess the practical value of this new approach, we have also performed traditional $G$-matrix calculations and compared the $V_{low-k}$ and $G$-matrix results between them and with experimental data. It turns out that  not only the former are in very good agreement with experiment, they are even somewhat better than the latter. 

 On the above grounds we may conclude that the $V_{low-k}$ approach provides an advantageous alternative to the usual $G$-matrix formalism in realistic shell-model calculations. We are currently using this approach for further nuclear structure studies. 
It is worth mentioning, however, that it is being profitably used also in other fields, as for instance the Fermi liquid theory for nuclear matter.\cite{brown01}

\section*{Acknowledgments}
This work was supported in part by the Italian Ministero dell'Universit\`a e della Ricerca Scientifica e Tecnologica (MURST) and by the U.S. DOE Grant No. DE-FG02-88ER40388.


\begin{thebibliography}{99}
\bibitem{sigma} We define 
$\sigma =\{(1/N_{d}) \sum_{i}[E_{exp}(i)-E_{calc}(i)]^{2}\}^{1/2}$, where
$N_{d}$ is the number of data.
\end{thebibliography}
\end{document}